\documentclass[conference,10pt]{IEEEtran}

\newcommand{\acro}{{\sf{HYDRA}}}
\newcommand{\sel}{{\sf{seL4}}}
\newcommand{\atp}{{\sf{$P_{Attest}$}}}
\newcommand{\att}{{\sf{Attest}}}

\newcommand{\vrf}{ {\ensuremath {\mathcal VRF}}}
\newcommand{\prv}{{\ensuremath {\mathcal PRV}}}
\newcommand{\key}{{\ensuremath {\mathcal K}}}

\usepackage{times}
\usepackage{subcaption}
\usepackage{subfiles,color}
\usepackage{float}
\usepackage[linesnumbered,ruled]{algorithm2e}
\usepackage[operators]{cryptocode}
\usepackage{booktabs}
\usepackage{multirow}
\usepackage{footnote}
\usepackage{listings}
\usepackage{authblk}
\begin{document}
\IEEEoverridecommandlockouts
\author{
	Karim Eldefrawy\textsuperscript{*}\thanks{\textsuperscript{*}Currently at the Computer Science Laboratory, SRI International. karim@csl.sri.com}\\
	Information and Systems Sciences Lab\\
	HRL Laboratories\\
	keldefrawy@hrl.com
	\and
	Norrathep Rattanavipanon\\
	Computer Science Department\\
	University of California, Irvine\\
	nrattana@uci.edu
	\and
	Gene Tsudik\\
	Computer Science Department\\
	University of California, Irvine\\
	gene.tsudik@uci.edu
}

\title{{\bf HYDRA}: {\bf HY}brid {\bf D}esign for {\bf R}emote {\bf A}ttestation \\ (Using a Formally Verified Microkernel)}

\newcommand{\karim}[1]{{\color{blue}[Karim: #1]}}
\newcommand{\oak}[1]{{\color{orange}[Oak: #1]}}
\newcommand{\gts}[1]{{\color{red}[Gene: #1]}}

\maketitle

\begin{abstract}
Remote Attestation (RA) allows a trusted entity 
({\em verifier}) to securely measure internal state of a remote untrusted 
hardware platform ({\em prover}). RA can be used to establish a static or 
dynamic root of trust in embedded and cyber-physical systems. It can also be used as a 
building block for other security services and primitives, such as software 
updates and patches, verifiable deletion and memory resetting. There are 
three major classes of RA designs: \emph{hardware-based}, \emph{software-based}, 
and \emph{hybrid}, each with its own set of benefits and drawbacks.

This paper presents the first hybrid RA design -- called \acro\ -- that builds upon 
formally verified software components that ensure memory isolation and 
protection, as well as enforce access control to memory and other resources. 
\acro\ obtains these properties by using the formally verified \sel\ microkernel.
(Until now, this was only attainable with purely hardware-based designs.) 
Using \sel\ requires fewer hardware modifications to the underlying 
microprocessor.
Building upon a formally verified software component increases confidence in security 
of the overall design of \acro\ and its implementation. We instantiate \acro\ on two commodity 
hardware platforms and assess the performance and overhead of 
performing RA on such platforms via experimentation; we show that \acro\ can attest 
$10$MB of memory in less than $500$msec when using a Speck-based message authentication code 
(MAC) to compute a cryptographic checksum over the memory to be attested.
\end{abstract}

\section{Introduction}
In recent years, embedded systems (ES), cyber-physical systems (CPS) and internet-of-things 
(IoT) devices, have percolated into many aspects of daily life, such as: households, 
offices, buildings, factories and vehicles. This trend of ``smart-ification" of everything that was previously
analog (or at least not connected) brings many obvious benefits. However, it also expands the
attack surface and turns these newly computerized gadgets into natural and attractive targets.

Remote Attestation (RA) is the process whereby a trusted entity called ``verifier" securely probes
internal state of a remote and untrusted hardware platform, called ``prover.'' 
RA can be used to establish a static or dynamic root of trust in ES, CPS and IoT devices.
Also, RA can be used as a foundation for constructing more specialized security services, e.g., 
software updates, verifiable deletion and memory resetting. There are three 
main classes of RA designs: \emph{hardware-based}, \emph{software-based}, and 
\emph{hybrid} (blending hardware and software). Each class has its own advantages
and limitations. This paper introduces \emph{the first} hybrid 
RA design -- called \acro\ -- based upon formally verified components 
to provide memory isolation and protection guarantees. Our main rationale is that designing RA 
techniques based upon such components increases confidence in security of such 
designs and their implementations. Of course, ideally, one would formally prove security of the entirety of 
an RA system, as opposed to proving security separately for each component and then proving that its 
composition is secure. However, we believe that this is not yet possible given the state-of-the-art in 
(automated) formal verification and synthesis of hardware and software.

One recent prominent example illustrating difficulty of correctly designing and implementing security 
primitives (especially, those blending software and hardware) is the TrustZone-based Qualcomm 
Secure Execution Environment (QSEE) kernel vulnerability and exploit reported in 
CVE-2015-6639 \cite{CVE-2015-6639}. ARM TrustZone \cite{TrustZone} 
is a popular System-on-Chip (SoC) and a CPU system-wide approach to security, it 
is adopted in billions of processors on various platforms.
CVE-2015-6639 enables privilege escalation and 
allows execution of code in the TrustZone kernel which can then be used 
to achieve undesired outcomes and expose keying material. This vulnerability 
was used to break Android's Full Disk Encryption (FDE) scheme 
by recovering the master keys \cite{Android-FDE-Break}. 
Our intention with this example is to demonstrate difficulty of getting both the design and the 
implementation right, as well as a motivation to use formally verified building blocks, which 
(we hope) will yield more secure RA techniques. To this end, our RA design uses the formally 
verified \sel\ microkernel to obtain memory isolation 
and access control. Such features have been previously attained with hardware in designs such as 
\cite{SMART} and \cite{trustlite}. Using \sel\ requires fewer hardware modifications to the 
underlying microprocessor and provides an automated formal proof of isolation guarantees of the 
implementation of the microkernel. \textit{To the best of our knowledge, this is the first attempt to design and implement 
RA using a formally verified microkernel.}

The main goal of this paper is to investigate  
a previously unexplored segment of the design space of hybrid RA schemes, specifically,  
techniques that incorporate formally verified and proven (using automated methods) components, such as the \sel\ microkernel. 
Beyond using \sel\ in our design, our implementation is also based on the formally verified 
executable of \sel; that executable is guaranteed to adhere to the formally verified and 
proven design. Another important goal, motivation and feature of our design is the expanded scope
of efficient hybrid RA techniques. While applicability of prominent prior results (particularly, 
SMART~\cite{SMART} and TrustLite~\cite{trustlite}) is limited to very simple 
single-process low-end devices, we target more capable devices that can 
run multiple processes and threads. 
We believe that this paper represents an important and necessary step towards building efficient
hybrid RA techniques upon solid and verified foundations. Admittedly, we do not verify our 
entire design and prove its security using formal methods. However, we achieve the next best 
thing by taking advantage of already-verified components and carefully arguing security of the overall design,
considering results on systematic analysis of features required for securely realizing hybrid RA \cite{DATE14}.
To achieve our goals we make two main contributions: (1) design of \acro\  -- the first hybrid RA technique based on 
the formally verified \sel\ microkernel which provides memory isolation and access control guarantees, (2) 
implementations of \acro\  on two commercially available development boards (Sabre Lite and ODROID-XU4) and 
their extensive analysis via experiments to demonstrate practicality of the proposed design. We show 
that \acro\ can attest $10$MB of memory in less than $500$ms when using Speck~\cite{simon-speck} as the underlying block-cipher
 to compute a cryptographic checksum over the memory to be attested.

\noindent {\em\bf Organization}: Section \ref{related-work} overviews related work, followed by 
Section \ref{goals-assumptions} which presents our goals and assumptions. The design of \acro\ is 
presented in Section \ref{designs} and its security analysis in Section \ref{security}. 
Implementation issues and performance assessment are discussed in Sections \ref{impl} and \ref{evaluation}. 


\section{Related Work \label{related-work}}
Prior work in remote attestation (RA) can be divided into 
three approaches: hardware-based, software-based, and hybrid.

\paragraph{Hardware-Based Remote Attestation}
The hardware-based approach typically relies on the security provided by 
a Trusted Platform Module (TPM) \cite{tpm}. A TPM is a secure co-processor designed to 
protect cryptographic keys, and utilize them to encrypt or digitally sign data. A TPM can also 
produce a summary (e.g., hash) of hardware and software configurations in the system.
A typical TPM also contains Platform Configuration Registers (PCR) that can be used 
as a secure storage of such a configuration summary. The values in PCRs can then 
be used as an evidence of attestation by accumulating an unforgeable chain of 
values of the system's state since the last reset. A TPM eventually signs these values with 
an attestation key along with a random challenge, provided by a verifier, and submits the 
computed result to the verifier. Gasmi et al. \cite{gasmi} presents how 
to link this evidence to secure channel end-points. 


\paragraph{Software-Based Remote Attestation}
Despite resisting all but physical attacks, the hardware-based 
approach is not suitable for embedded devices due to its 
additional software complexity and expense.
Therefore, many software-only RA approaches have 
been proposed, specifically for embedded devices. Pioneer \cite{pioneer} is among the 
first to study RA without relying on any secure co-processor or 
CPU-architecture extensions. The main idea behind Pioneer is to create a special checksum function with 
run-time side-effects (e.g., status registers) for attestation. Any malicious emulation of 
said checksum function can be detected through additional timing overhead incurred from the 
absence of those side-effects. 
Security of this approach became 
questionable after several attacks on such schemes (i.e., \cite{Castelluccia}) were 
demonstrated.

\paragraph{Hybrid Remote Attestation}
The main shortcoming of the software-based approach is 
that it makes strong assumptions about adversarial capabilities, which 
may be unsatisfied in practical networked settings \cite{DAC16-1}. Thus, several hybrid 
software-hardware co-designs have been proposed 
to overcome this limitation. 
SMART \cite{SMART} presents a hybrid approach for RA with minimal 
hardware modifications to existing MCUs. In addition to having uninterruptable 
attestation code and attestation keys resided in ROM, this architecture utilizes a 
hard-wired memory protection unit (MPU) to restrict access to secret keys to only 
SMART code. The attestation is performed inside ROM-resident attestation code 
by computing a cryptographic checksum over a memory region and returning the 
value to the verifier. 
TrustLite \cite{trustlite} extends \cite{SMART} to enable RA while
supporting an interrupt handling in a secure place.



In addition to the above work designing RA schemes, 
\cite{DATE14} provides a systematic treatment of RA by presenting 
a precise definition of the desired service and
proceeding to its systematic deconstruction into necessary and
sufficient (security) properties. These properties are then mapped into
a minimal collection of hardware and software components that
results in secure RA. We build on the analysis in \cite{DATE14}  and utilize these 
properties and components (which are described in Section \ref{goals-assumptions}) 
and show how to instantiate them in new ways to develop our new 
hybrid RA design, \acro.

\section{Goals and Assumptions \label{goals-assumptions}}
This section presents an overview of \acro\ and the rationale behind its design, 
discusses the common remote attestation (RA) security objective and features, and the adversary model. 

\subsection{Design Rationale}
Our main objective is to explore a new segment of the 
design space for RA schemes. Our hybrid RA design -- \acro\ --
requires very little in terms of secure hardware and builds upon the formally verified
\sel\ microkernel.  As shown in Section \ref{impl}, hardware support needed by \acro\ is 
readily available on commercial off-the-shelf development boards and processors, 
e.g., Sabre Lite and ODROID-XU4 boards. The rationale behind our design is that \sel\ offers
certain guarantees (mainly process isolation and access control to memory and resources) that 
provide RA features that were previously feasible only using hardware components.
In particular, what was earlier attained 
using additional MCU controls and Read-Only Memory (ROM) in the 
SMART~\cite{SMART} and TrustLite~\cite{trustlite} architectures can now be instantiated 
using capability controls in \sel.

To motivate and justify the design of \acro, we start from the work by Francillon, et al. \cite{DATE14}. It
provides a systematic treatment of RA by developing a semi-formal definition of RA as a 
distinct security service, and systematically deconstructing it into a necessary and
sufficient security objective, from which specific properties are derived. 
These properties are then mapped into a collection of hardware and software 
components that results in an overall secure RA design. Below, we summarize the security 
objective in RA and its derived security properties. In Sections \ref{designs} and \ref{impl}
we show how the security objective and each property 
are satisfied in \acro\ design and instantiated in two concrete prototypes 
based on the Sabre Lite and ODROID-XU4 boards.


%

\subsection{Hybrid RA Security Objective and Derived Properties}
According to \cite{DATE14}, the RA security objective is
to allow a (remote) prover ($\mathcal{PRV}$) to create an unforgeable 
authentication token, that convinces a verifier ($\mathcal{VRF}$) that the former is in
some well-defined (expected) state. Whereas, if $\mathcal{PRV}$ has been
compromised and its state has been modified, the authentication
token must reflect this. \cite{DATE14} also describes a 
combination of platform properties that achieve aforementioned security objective. 
The goal of the analysis in \cite{DATE14} is to obtain 
a set of properties that are both necessary and sufficient for secure RA. 
The security objective is examined and the properties 
needed to attain it are identified. The conclusion is that  
the following properties collectively represent the minimal and necessary set to 
achieve secure RA on any platform.
\begin{itemize}
\item \textit{Exclusive Access to Attestation Key (\key):}  the attestation 
process (\atp) must be the only one with access to \key. This 
is the most difficult property to impose on embedded (especially, low-end and mid-range) devices. As 
argued in  \cite{DATE14}, this property is unachievable without some hardware support on 
low-end devices. If the underlying processor supports multiple privilege modes and a full-blown
separation of memory for each process, one could use a privileged process to handle all computations 
that involve the key. However, low-end and medium-end processors generally do not offer such ``luxury''
features. 
\item \textit{No Leaks:} no information related to (or derived from) \key\ must be accessible 
after the execution of \atp\ completes. To  achieve this,  all intermediate values that depend on the key -- except 
the final attestation token to be returned to $\mathcal{VRF}$ -- must be securely erased, when they are 
no longer needed. This is again applicable to very low-end devices, with none or minimal Operating System (OS) support and
assuming that memory is shared between processes. However, if the underlying hardware and/or software 
(i.e., OS) guarantees that each process' memory is inaccessible by any other process, then this property is 
trivially satisfied.
\item \textit{Immutability:}  To ensure that the attestation executable (\att) cannot be modified,
SMART \cite{SMART} and the design in \cite{DATE14} place it in ROM, which is available on most, even low-end, platforms. 
ROM is certainly one relatively inexpensive way to enforce \att's code immutability. However, owing to
\sel's features, \att\ in \acro\ is resident in, and executed from, RAM.          
\item \textit{Uninterruptability:}  Execution of \att\ must be uninterruptible. This is necessary to 
ensure that malware does not obtain the key (or some function thereof) by interrupting \att\ while
any key-related values remain in registers or other accessible locations. As discussed later, \acro\ achieves this
property by assuring that it runs with the highest possible priority under \sel. 
%
\item \textit{Controlled Invocation (aka Atomicity):}  \att\ must only be invocable from its first instruction 
and must exit only at one of its legitimate last (exit) instruction. This is motivated by the need 
to prevent code-reuse attacks. 
%

       
%
\end{itemize}
\cite{DATE14} has one additional property of \textit{Secure Reset} when 
someone attempts to start the attestation executable (\att) from the middle. We argue that a complete secure 
reset is not needed if controlled invocation is enforced. All that is needed in that case 
is to raise an exception as long as the memory space of \att\ is 
protected and integrity of executable is guaranteed.

\subsection{Adversarial Model \& Other Assumptions}
Based on the recent taxonomy in \cite{DAC16-1}, adversaries 
in the RA context can be categorized as follows:
\begin{itemize}
\item \textit{Remote Adversaries:} exploit vulnerabilities in
a prover's software to inject (over the network) malware onto it.
\item \textit{Local Adversaries:} locate sufficiently near in order to
eavesdrop on, and manipulate, the prover's communication channel(s).
\item \textit{Physical Adversaries:} have full (local) physical access to the prover and its hardware
and can perform physical attacks, e.g., side-channel attacks to obtain keys,
physically extract memory values, and modify the states of various hardware components.
\end{itemize}
Similar to prior hybrid RA designs, \acro\ aims to protect against remote and local
adversaries, while physical adversaries are out-of-scope. We note that,
at least in a single-prover setting\footnote{See \cite{darpa} for physical attack
resilience in groups or swarms of provers.}, protection against physical attacks
can be attained by encasing the CPU in tamper-resistant coating
and employing standard techniques to prevent side-channel key leakage. 
These include: anomaly detection, internal power regulators and
additional metal layers for tamper detection. 
We consider $\mathcal{PRV}$ to be a (possibly) unattended
remote hardware platform running multiple processes 
on top of \sel.  Once $\mathcal{PRV}$ 
boots up and runs in steady state, the adversary ($\mathcal{ADV}$) might exercise 
complete control over all application software (including code and data) 
before and after execution of \atp. 
As physical attacks are out of scope, $\mathcal{ADV}$ cannot 
induce hardware faults or retrieve
a stored \key\ using side channels. 
$\mathcal{ADV}$ also has no
means of interrupting execution of \sel\ or the \atp\ code when invoked (how this is ensured will be demonstrated later). 
Finally, recall that $\mathcal{PRV}$ and $\mathcal{VRF}$ must share at least one
secret (attestation) key $\mathcal{K}$. This key can be pre-loaded onto $\mathcal{PRV}$ at
installation time; it is stored as part of \atp\ binaries. 
We do not address the details of this procedure.

\section{HYDRA Design\label{designs}} 
We first overview \sel\ and discuss how it can be used in \acro\ 
to realize previously identified RA properties. We then describe the sequence of operations in \acro. 

\begin{figure}[!t]
\centering
\includegraphics[width=\linewidth]{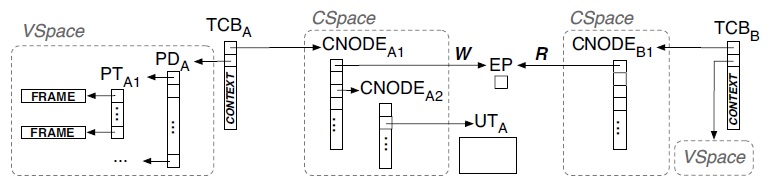}
\caption{Sample \sel\ instantiation from \cite{sewell2011sel4}.}
\label{sel4-example}
\end{figure}

\subsection{\sel\ Overview }
\sel\ is a member of the L4 microkernel family, specifically designed 
for high-assurance applications by providing isolation and memory 
protection between different processes. These 
properties are mathematically guaranteed by a full-code level functional correctness proof,
using automated tools. A further correctness proof of the C code translation 
is presented in \cite{sel4-binary-proof}, thus extending functional 
correctness properties to the binary level without needing a trusted compiler.
Therefore, behavior of the \sel\ binary strictly adheres to, and is fully 
captured by, the abstract specifications.

Similar to other operating systems, \sel\ divides the virtual memory
into two separated address spaces: \emph{kernel-space} and \emph{user-space}. 
The kernel-space is reserved for the execution of the \sel\ microkernel while
the application software is run in user-space.
By design, and adhering to the nature of microkernels, the \sel\ microkernel provides minimal 
functionalities to user-space applications: thread, inter-process 
communication (IPC), virtual memory, capability-based access control
and interrupt control. The \sel\ microkernel leaves the implementations of other traditional
operating system functions -- such as device drivers and file systems -- to user-space. 

Figure \ref{sel4-example} (borrowed from \cite{sewell2011sel4}) 
shows an example of \sel\ instantiation with two 
threads -- sender A and receiver B  -- that communicate via an \emph{EndPoint} EP. 
Each thread has a \emph{Thread Control Block} (TCB) that 
stores its context, including: stack pointer, program counter, register values, as 
well as pointers to \emph{Virtual-address Space} (VSpace) and \emph{Capability Space} 
(CSpace). VSpace represents available memory regions
that the \sel\ microkernel allocated to each thread. The root of VSpace represents 
a \emph{Page Directory} (PD), which contains \emph{Page Table} (PT) 
objects. \emph{Frame} object representing a region of physical memory resides
in a PT. Each thread also has its own kernel managed CSpace
used to store a \emph{Capability Node} (CNode) and \emph{capabilities}. CNode is a 
table of slots, where each slot represents either a capability or another CNode.

A capability is an unforgeable token representing an access control authorization of each kernel object or component.
A thread cannot directly access or modify a capability since CSpace is managed by, and stored inside, the kernel.
Instead, a thread can invoke an operation on a kernel object by providing a pointer to a capability
that has sufficient authority for that object to the kernel. For example, sender A in Figure \ref{sel4-example} 
needs a write capability of EP for sending a message, while receiver B needs a read capability to receive a message.
Besides read and write, \emph{grant} is another access right in \sel, available only for an endpoint object.
Given possession of a grant capability for an endpoint, any capability from the possessor 
can be transferred across that endpoint. For instance, if A in Figure \ref{sel4-example} has grant access to EP, it
can issue one of its capabilities, say a frame, to B via EP. Also, capabilities can be 
statically issued during a thread's initialization by the {\em initial process}. 
The initial process is the first executable user-space process loaded into working memory (i.e., RAM) after the \sel\ 
microkernel is loaded. This special process then forks all other processes.
Section \ref{subsec:sop} describes the role, the details and the capabilities of the initial process 
in \acro\ design.

\sel's main ``claim to fame"  is in being the first formally verified general-purpose operating system.
Formal verification of the \sel\ microkernel is performed by interactive, machine-assisted and machine-checked proof
using a theorem prover Isabelle/HOL. Overall functional correctness is obtained 
through a \emph{refinement} proof technique, which demonstrates that the binary of \sel\ 
refines an abstract specification through three layers of refinement. Consequently (under some 
reasonable assumptions listed in Appendix \ref{appdx:sel4_assump}) the \sel\ binary is fully captured by the 
abstract specifications. In particular, two important feature derived from \sel's abstract specifications, are that: 
{\bf the kernel never crashes}. Another one  is that: {\bf every kernel API call always terminates and returns to 
user-space}. Comprehensive details of \sel's formal verification can be found in \cite{klein2014comprehensive}.

Another \sel\ feature very relevant to our work is: \textbf{correctness of access control enforcement} 
derived from functional correctness proof of \sel. \cite{sewell2011sel4} and \cite{murray2013sel4} 
introduce formal definitions of the access control model and information flow in \sel\ at the 
abstract specifications. They demonstrate 
the refinement proof from these modified abstract specifications to the C implementation using
Isabelle/HOL theorem prover, which is later linked to the binary level (by the same theorem prover).
As a result, three properties are guaranteed by the access control enforcement proof: (1) 
\emph{Authority Confinement}, (2) \emph{Integrity} and (3) \emph{Confidentiality}. Authority 
confinement means that authority propagates correctly with respect to its capability. For example, 
a thread with a read-only capability for an object can only read, and not write to, that object. Integrity 
implies that system state cannot be modified without explicit authorization. For instance, a read 
capability should not modify internal system state, while write capability should only modify an 
object associated with that capability. Finally, confidentiality means that an object cannot be read or 
inferred without a read capability. Thus, the proof indicates that access control in \sel, once specified 
at the binary level, is correctly enforced as long as the \sel\ kernel is active. 

We now show how \sel's access control enforcement property satisfies required RA features.

\begin{table*}
\centering
\caption{Security Properties in Hybrid RA Designs}
\label{table:features}
\begin{tabular}{lccc}
\toprule 

	\multirow{2}[3]{*}{Security Property} & \multirow{2}[3]{*}{SMART \cite{SMART}} & \multirow{2}[3]{*}{TrustLite \cite{trustlite}}  &\multirow{2}[3]{*}{\acro}	  \\
	\\
    	\midrule \midrule
	Exclusive Access to \key & HW (Mod. Data Bus) & SW (programmed MPU) & SW (\sel) \\
	\hline \\[-.5em]
	No Leaks & SW (CQUAL and Deputy)  & HW (CPU Exception Engine) & SW (\sel)\\
	\hline \\[-.5em]
	Immutability & HW (ROM) & HW (ROM) and SW (programmed MPU) & HW (ROM) and SW (\sel)\\
	\hline \\[-.5em]
	Uninterruptability & SW (Interrupt Disabled) & HW (CPU Exception Engine) & SW (\sel) \\
	\hline \\[-.5em]
	Controlled Invocation & HW (ROM) & HW (ROM) &  SW (\sel) \\
    	\bottomrule
\end{tabular}
\end{table*}

\subsection{Deriving \sel\ Access Controls from Properties}
We now describe access control configuration of \sel\ user-space that achieves most 
required properties for secure RA, as described in section \ref{goals-assumptions}. 
We examine each feature and identify the corresponding access control configuration.
Unlike previous hybrid designs, \acro\ pushes almost all of these required features
into software, as long as the \sel\ microkernel boots correctly. (A comparison with SMART and TrustLite is 
in Table \ref{table:features}.) 

\paragraph{Exclusive Access to \key}
directly translated to an access control configuration.
Similar to previous hybrid approaches, \key\ can be hard-coded into the 
\att\ at production time.
Thus, \att\ needs to be configured to be accessible only to \atp.

\paragraph{No Leaks}
achieved by the separation of virtual address space.
Specifically, the virtual memory used for \key-related computation 
needs to be configured to be accessible to only \atp.

\paragraph{Immutability}
achieved using combination of verifiable boot and runtime isolation guarantee from \sel.
At runtime, \att\ must be immutable, which can be guaranteed
by restricting the access control to the executable to only \atp.
However, this is not enough to assure immutability of \att\ executable because \att\ can
be modified after loaded into RAM but before executed. Hence, a verifiable boot of \att\
is required.

\paragraph{Uninterruptability}
satisfied by setting the scheduling priority of \atp\ higher
than other processes since the formal proof of \sel\ scheduling mechanism guarantees that a lower priority process
cannot preempt the execution of a higher priority process. In addition, \sel\ guarantees that, once set,
 the scheduling priority in \sel\ cannot be increased at runtime (but possible to decrease the priority
value).

Note that this feature implies \atp\ needs to be the initial user-space process 
since the \sel\ microkernel always assigns the highest priority to the initial process.

\paragraph{Controlled Invocation}
achieved by the isolation of process' execution.
In particular, TCB of \atp\ cannot be accessed or manipulated by other processes.


With these features, we conclude that the access control configuration of \sel\ user-space
needs to (at least) include the following:
\begin{itemize}
\item[C1] \atp\ has exclusive access to \att; this also includes \key\ residing in \att.
(Recall that \atp\ is the attestation {\bf process}, while \att\ is the executable that actually performs attestation.)
\item[C2] \atp\ has exclusive access to its TCB.
\item[C3] \atp\ has exclusive access to its VSpace.
\end{itemize}
Even though this access control configuration can be enforced at the binary code level,
this assumption is based on that \sel\ is loaded into RAM correctly. However, this can
be exploited by an adversary by tricking the boot-loader to boot his malicious \sel\ microkernel
instead of the formally verified version and insert a new configuration violating above access controls. 
Thus, the hardware signature check of the \sel\
microkernel code is required at boot time. The similar argument can also be made for
\atp' code. As a result, additional integrity check of \atp'
code needs to be performed by \sel\ before executing.

\subsection{Building Blocks}

In order to achieve all security properties described above, 
\acro\ requires the following four components.
\paragraph{Read-Only Memory}
region primarily storing immutable data (e.g. hash of 
public keys or signature of software) required for secure boot of the \sel\ microkernel.

\paragraph{MCU Access Control Emulation}	
high-assurance software framework capable of emulating MCU access controls
to attestation key \key. At present, \sel\ is the only formally 
verified and mathematically proven microkernel capable of this task.

\paragraph{Attestation Algorithm}
software residing in \atp\ and 
serving two main purposes: authenticating an attestation request, 
and performing attestation on memory regions.

\paragraph{Real-Time Clock}
loosely synchronized (with \vrf) real-time clock for detecting replayed, reordered 
and delayed requests. If \prv\ does not have a clock, 
a secure counter can replace a real-time clock with a sacrifice of the 
delayed message detection. 

\begin{figure}[!t]
\centering
\includegraphics[width=6cm]{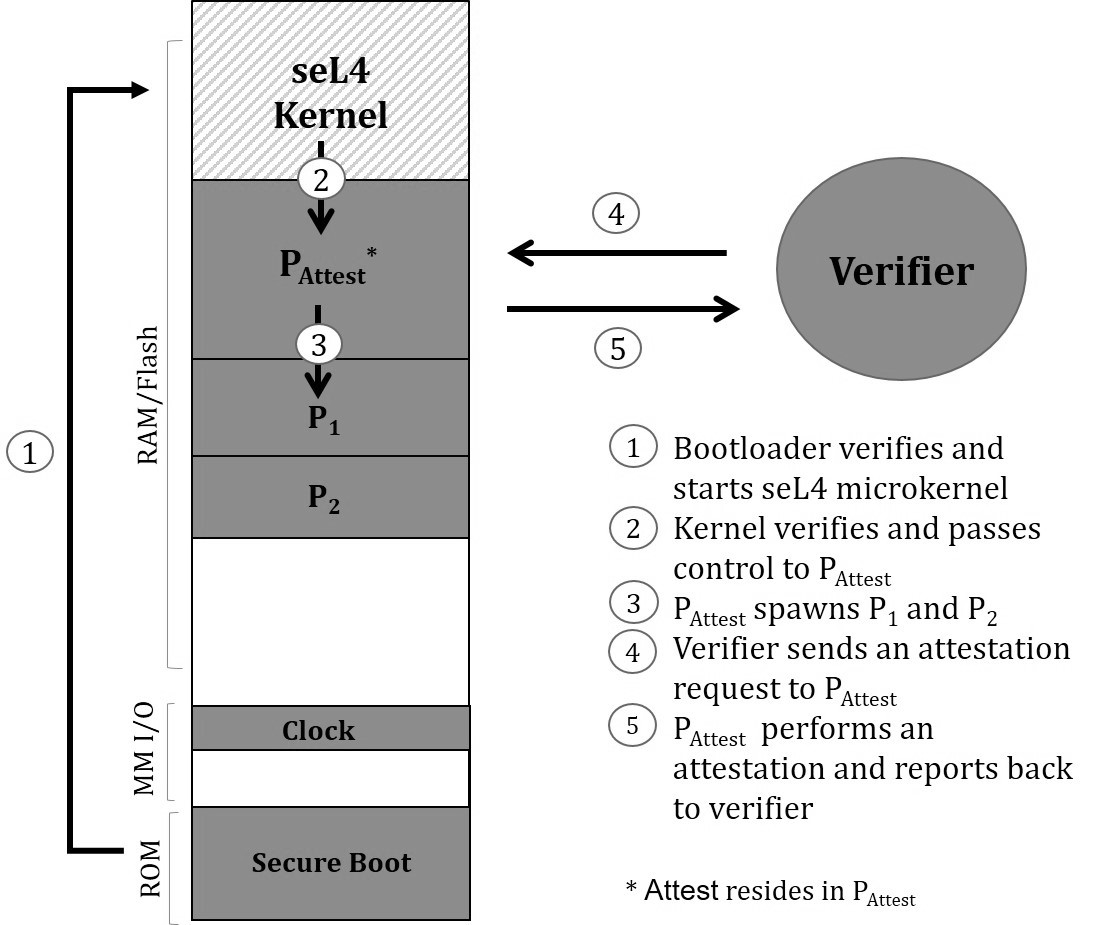}
\caption{Sequence of Operation in \acro}
\label{sop}
\end{figure}

\subsection{Sequence of Operation \label{subsec:sop}}
The sequence of operations in \acro, shown in Figure \ref{sop}, has three steps: 
boot, setup, and attestation. 

\paragraph{Step 1: Boot Process}
Upon a boot, \prv\ first executes a ROM-resident boot-loader. The boot-loader verifies authenticity and 
integrity of the \sel\ microkernel binary. Assuming this verification succeeds, the boot-loader loads all executables, 
including kernel and user-space, into RAM and hands over control to the \sel\ microkernel. Further details of secure 
boot in our prototype can be found in Section \ref{impl}.

\paragraph{Step 2: \sel\ Setup}
The first task in this step is to have the \sel\ microkernel setting up the user-space and then starting \atp\ as 
the initial user-space process. Once the initialization inside the kernel is over, the \sel\ microkernel gathers 
capabilities for all available memory-mapped locations and assigns them to \atp. The \sel\ kernel also performs 
an authenticity and integrity check of \atp\ to make sure that it has not been modified. After successful 
authentication, the \sel\ microkernel passes control to \atp.

With full control over the system, \atp\ starts the rest of user-space with a lower scheduling priority and 
distributes capabilities that do not violate the configuration specified earlier.
After completing configuration of memory capabilities and starting the rest of the user-space, 
\atp\ initializes the network interface and waits for an attestation request.

\paragraph{Step 3: Attestation}
An attestation request, sent by a verifier, consists of 4 parameters: 
(1) $T_R$ reflecting \prv's time when the request was generated,
(2) target process $p$, (3) its memory range $[a, b]$ that needs to be attested,
and (4) cryptographic checksum $C_R$ of the entire attestation request.

Similar to SMART~\cite{SMART}, the  cryptographic checksum function used in 
attestation is implemented as a Message Authentication Code (MAC), 
to ensure authenticity and integrity of attestation protocol messages.

Upon receiving an attestation request \atp\ checks whether $T_R$ is within an acceptable range
of the \prv's real-time clock before performing any cryptographic operation; this is in order to mitigate 
potential DoS attacks. If $T_R$ is not fresh, \atp\ ignores the request and returns to the waiting state.
Otherwise, it verifies $C_R$. If this fails, \atp\ also abandons the request and returns to the waiting state.

Once the attestation request is authenticated, \atp\ computes a cryptographic checksum of the memory region 
[$a$, $b$] of process $p$.
Finally, \atp\ returns the output to \vrf. The pseudo-code of this process is shown in 
Algorithm \ref{alg:mod_smart}.	

\begin{algorithm}
	\small
	\SetKwInOut{Input}{Input}
	\SetKwInOut{Output}{Output}

	\Input{$T_R$ timestamp of request 
			\\ $p$ target process for attestation 
			\\ $a, b$ start/end memory region of target process
			\\ $C_R$ cryptographic checksum of request}
    	\Output{Attestation Report}

	\Begin {

		/* Check freshness of timestamp and verify request */

		\If{$\neg$ CheckFreshness($T_R$)} {exit()\;}

		\If{$\neg$ VerifyRequest($C_R$, $K_{Auth}$, $T_R$\concat$p$\concat$a$\concat$b$)} {exit()\;}

		/* Retrieve address space of process $p$ */

		$Mem$ $\leftarrow$ RetrieveMemory($p$);

		/* Compute attestation report  */

		MacInit(\key);

		MacUpdate($T_R$\concat$p$\concat$a$\concat$b$)\;

		\For{$i \in [a, b]$}{MacUpdate($Mem$[$i$]);}

		$out$ $\leftarrow$ MacFinal()\;
		
		\Return {$out$}
	}
	\caption{\att\  Pseudo-Code}
	\label{alg:mod_smart}

\end{algorithm}

\section{HYDRA Implementation\label{impl}} 
To demonstrate feasibility and assess practicality of \acro, we 
implement two prototypes of it on two commercially available hardware 
platforms: ODROID-XU4 \cite{odroid} and Sabre Lite
\cite{sabre}. We focus here more on the Sabre Lite 
implementation due to the lack of an \sel~compatible 
networking (e.g., Ethernet) driver and a programmable ROM in current 
ODROID-XU4 boards. 
Section \ref{evaluation} presents a detailed performance evaluation of the 
\acro~implementation.

\subsection{\sel~User-space Implementation}

Our prototype is implemented on top of version 1.3 of the \sel\ microkernel~\cite{seL4-website}.
The complete implementation, including helper libraries and the networking stack,
consists of $105,360$ lines of C code (see Table \ref{loc-table} for a more 
detailed breakdown). The overall size of executable is $817$KB whereas the base \sel\ microkernel size is $215$KB.
Excluding all helper libraries, the implementation of \acro\ is just $600$ lines of C code.
In the user-space, we base our C code on following libraries: seL4utils, seL4vka and seL4vspace; 
these libraries provide the abstraction of processes, memory management and virtual space respectively.
In our prototypes, \atp\ is the initial process in the user-space and receives capabilities 
to all memory locations not used by \sel. Other processes in user-space are
spawned by this \atp. We also ensure that access control of those processes
does not conflict with what we specified in Section \ref{designs}. The details of this access control
implementation are described below in this section.
 
The basic C function calls are implemented in muslc library.
seL4bench library is used to evaluate timing and performance of our \acro\ implementation. 
For a timer driver, we rely on its implementation in seL4platsupport.
All source code for these helper libraries can be found in \cite{sel4-libs} and
these libraries contribute around $50$\% of the code base in our implementation.
We use an open-source implementation of a network stack and an 
Ethernet driver in the user-space \cite{sel4-network-stack}.
We argue that this component, even though not formally verified, should not 
affect security objective of \acro\ as long as an IO-MMU is used to restrict 
Direct Memory Access (DMA) of an Ethernet driver. 
The worst case that can happen from not formally verified network stack 
is symmetrical denial-of-service, which is out of scope of \acro.

\begin{table}
\centering
\caption{Complexity of HYDRA Impl. on Our Prototype}
\label{loc-table}
\resizebox{\columnwidth}{!}{%
\begin{tabular}{ccccc}
\toprule 

	 \multirow{2}{*}{Complexity} & HYDRA with & HYDRA w/o  & HYDRA w/o & \sel ~Kernel  \\
    	&  net. and libs &net. stack & net. and libs & Only  \\
    	\midrule
	LoC & 105,360 & 68,490 & 11,938 & 9,142 \\
	Exec Size & 817KB & 721KB & N/A & 215KB \\
    	\bottomrule
\end{tabular}
}
\end{table}

\begin{figure}[!htp]
\centering
\includegraphics[width=2in]{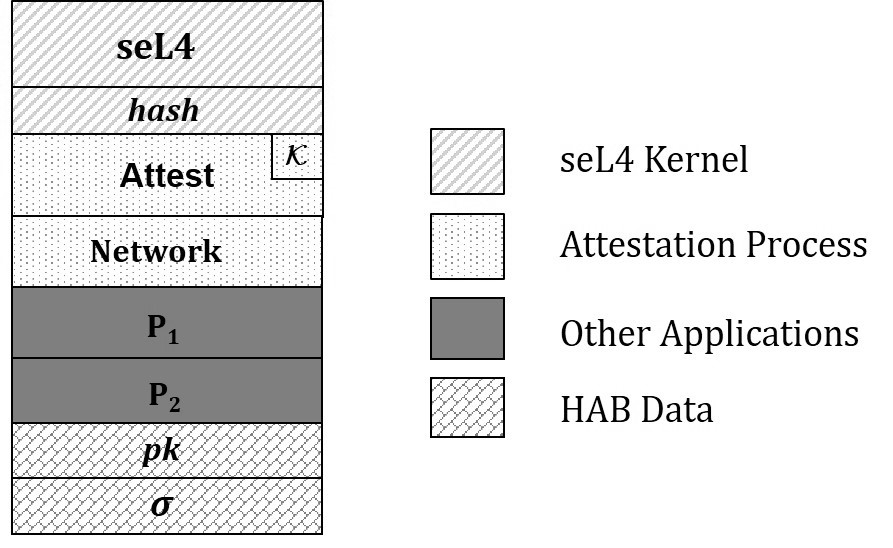}
\caption{Image Layout in Flash}
\label{img_layout}
\end{figure}

\subsection{Secure Boot Implementation}
 
Here, we describe how we integrate an existing secure boot feature (in Sabre Lite) with 
our \acro\ implementation.

\paragraph{Secure Boot in Sabre Lite}
NXP provides a secure boot feature for Sabre Lite boards, called High Assurance Boot (HAB) \cite{hab}.
HAB is implemented based on a digital signature scheme with public- and private-keys.
A private-key is required to generate a signature of the software image during manufacturing whereas 
a public-key is used by ROM APIs for decrypting and verifying the software signature at boot time.
A public-key and a signature are attached to the end of software image, which is pre-installed 
into a flash medium during manufacturing. The digest of a public-key is fused into a one-time 
programmable ROM in order to ensure the authenticity of the public-key 
and the booting software image.
At boot time, the ROM boot-loader first loads the software image into RAM
and then verifies an attached public-key by comparing it with the 
reference hash value in ROM. It then authenticates the software image through the attached 
signature and the verified public-key. Execution of this image is allowed only if signature verification succeeds.
Without a private-key, an adversary cannot forge a legitimate digital signature and thus is 
unable to insert and boot his malicious image. 

\begin{figure*}[!t]
\centering
\includegraphics[width=6in]{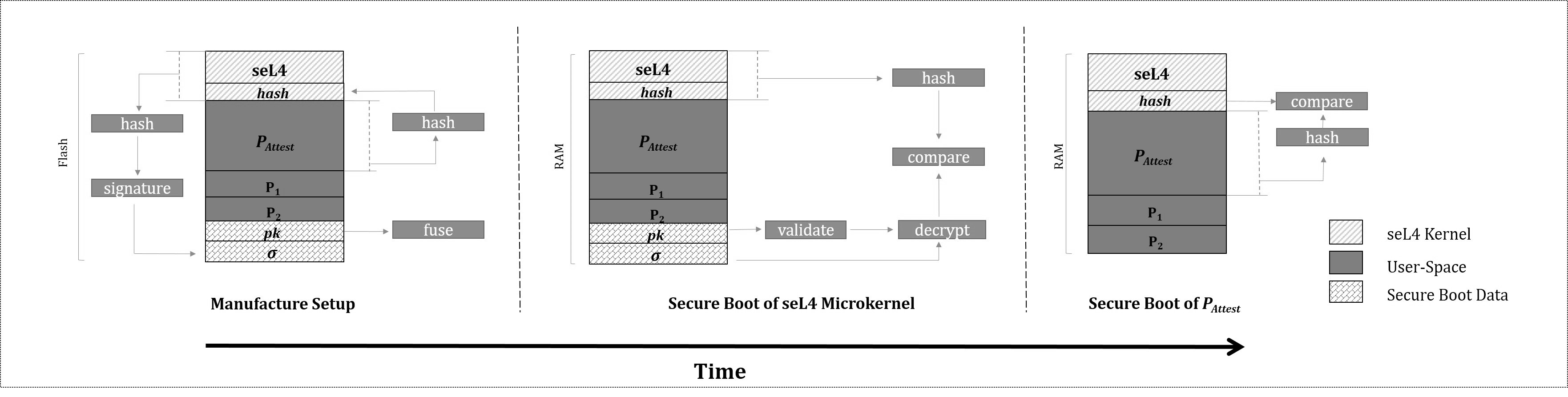}
\caption{Secure Boot Sequence in Sabre Lite Prototype}
\label{sb-seq}
\end{figure*}

\paragraph{Secure Boot of \acro}
HAB can be utilized to ensure that the \sel\ microkernel is the first program 
initialized after the ROM boot-loader. Therefore, the entire \sel\ microkernel binary 
code can be covered when computing the digital signature by HAB during 
manufacturing. Moreover, the \sel\ microkernel needs to be assured that 
it gives control of the user-space to the verified \atp. Thus, the \sel\
microkernel has to perform an integrity check of \atp\ 
before launching it. As a result, a hash of \att\ 
needs to be included in the \sel\ microkernel's binaries
during production time and be validated upon starting the initial process.

With this procedure, a chain of trust is established in the remote attestation 
system in \acro. This implies that no other programs, except the \sel\ 
microkernel can be started by the ROM boot-loader and consequently only 
\atp\ is the certified initial process in the user-space, which
achieve the goal of secure boot of remote attestation system.
Figure \ref{sb-seq} illustrates the secure boot of \acro\ in Sabre Lite prototype.


\subsection{Access Control Implementation}

Here we describe how the access control configuration specified in 
section \ref{designs} is implemented in our \acro\ prototype. Our goal 
is to show that in the implementation of \acro\ no other user-space 
processes, except \atp,
can have any kind of access to: (1) the binary executable code (including \key), (2) the virtual 
address space of \atp, and (3) the TCB of \atp.
To provide those access restrictions in the user-space, we make sure that 
we do not assign capabilities associated to those memory regions to other 
user-space processes.
Recall that \atp\ as the initial process contains all capabilities 
to every memory location not used by the \sel\ microkernel. And there are two ways 
for \atp\ to issue capabilities: 
dynamically transfer via endpoint with grant access right or statically 
assign during bootstrapping a new process.

In our implementation, \atp\ does not create any 
endpoint with grant access, which disallows any capability of 
\atp\ to transfer to a new process after created.
Thus, the only way that capabilities can be assigned to a new process
is before that process is spawned. When creating a new process, 
\atp\ assigns only minimal amount of capabilities 
required to operate that process, e.g. in our prototype, only the 
CSpace root node and fault endpoint (used for receiving IPCs 
when this thread faults) capabilities are assigned to any newly 
created process. Limited to only those capabilities, any 
other process cannot access the binary executable code as well 
as existing virtual memory and TCB of \atp.

Moreover, during bootstrapping the new process, \atp\ 
creates a new PD object serving as 
the root of VSpace in the new process. This is to ensure that any 
new process' virtual address space is initially empty and does not 
overlap with the existing virtual memory of \atp. 
Without any further dynamic capability distribution, this 
guarantees that other processes cannot access any memory 
page being used by \atp.
Sample code for configuring a new process in our prototype
is provided in Appendix \ref{appdx:new_process_code}.


\subsection{Key Storage}
Traditionally, in previous hybrid designs, a prover device requires a special hardware-controlled memory location
for securely storing $\mathcal{K}$ and protecting it from software attacks.
However, in \acro, it is possible to store $\mathcal{K}$ in a normal memory location (e.g. flash) due to the formally 
verified access control and isolation properties of \sel.
Moreover, since $\mathcal{K}$ is stored in a writable memory, its update can easily happen without any secure hardware involvement.
Thus, in our prototypes, $\mathcal{K}$ is hard-coded at production time and stored in the same region as \att.



Besides $\mathcal{K}$, \acro\ contains another key, $K_{Auth}$, used for verifying an authenticity of an attestation request.
$K_{Auth}$ can be a separate key and pre-stored next to \key\ during a production time;
or, $K_{Auth}$ can be derived from $\mathcal{K}$ through a key derivation function (KDF) at runtime as well.

\subsection{Timestamp Generation}
Recall (Section \ref{designs}) that a timestamp generated by a loosely synchronous 
real-time clock is required for ensuring freshness of attestation requests. 
There is currently no implementation of drivers for real-time clock in \sel. 
We generate a pseudo-timestamp by a counter, whose driver 
is provided by seL4platsupport, and a timestamp of the first validated request, as follows: When 
a device first wakes up and securely start \atp.
\atp\ promptly loads a timestamp, $T_{save}$, that was previously saved
in a separated location of flash medium before the last reset. When the first attestation request arrives, 
\atp\ checks its attached timestamp, $T_{first}$, whether it is greater than $T_{save}$
and proceeds to $VerifyRequest$ if that is the case. After the assurance of the validated request, 
\atp\ keeps track of $T_{first}$ and start a counter. At any time afterwards, a timestamp can be 
constructed by combining the current counter value with $T_{first}$. In addition, \atp\ periodically generates 
and saves this timestamp value into flash medium for the next reboot's usage.

\section{Security Analysis\label{security}}
\begin{figure*}[!h]
	
 \centering
    	\begin{subfigure}[t]{0.3\textwidth}
		\centering
		\includegraphics[width=\linewidth]{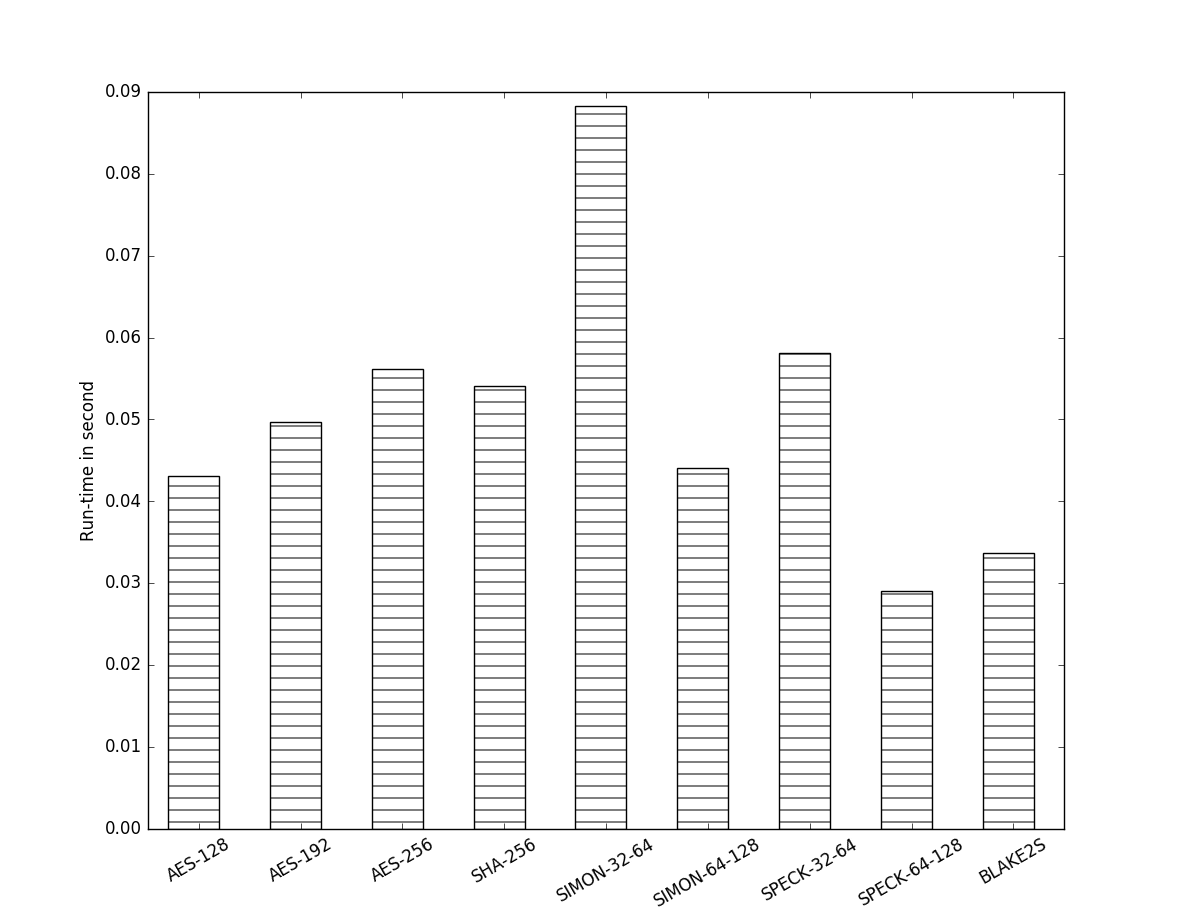}
		\caption{MAC Implementations}
		\label{macfn}
	\end{subfigure}
	~
    	\begin{subfigure}[t]{0.3\textwidth}
		\centering
		\includegraphics[width=\linewidth]{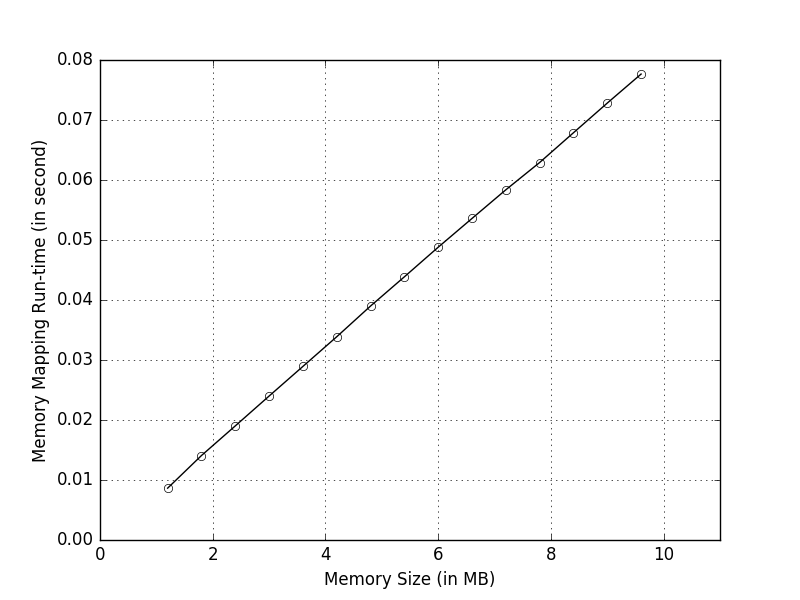}
		\caption{Memory Mapping in \sel}
		\label{retrieve_mem}
	\end{subfigure}
	~
	\begin{subfigure}[t]{0.3\textwidth}
		\centering
		\includegraphics[width=\linewidth]{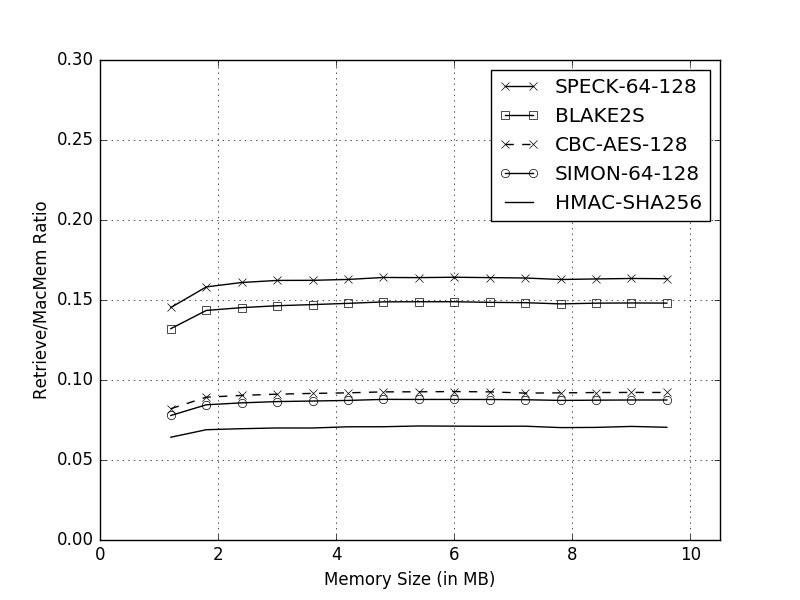}
		\caption{MacMem and RetrieveMem}
		\label{retr_mac_ratio}
	\end{subfigure}
	~
	\medskip
	\begin{subfigure}[t]{0.3\textwidth}
		\centering
		\includegraphics[width=\linewidth]{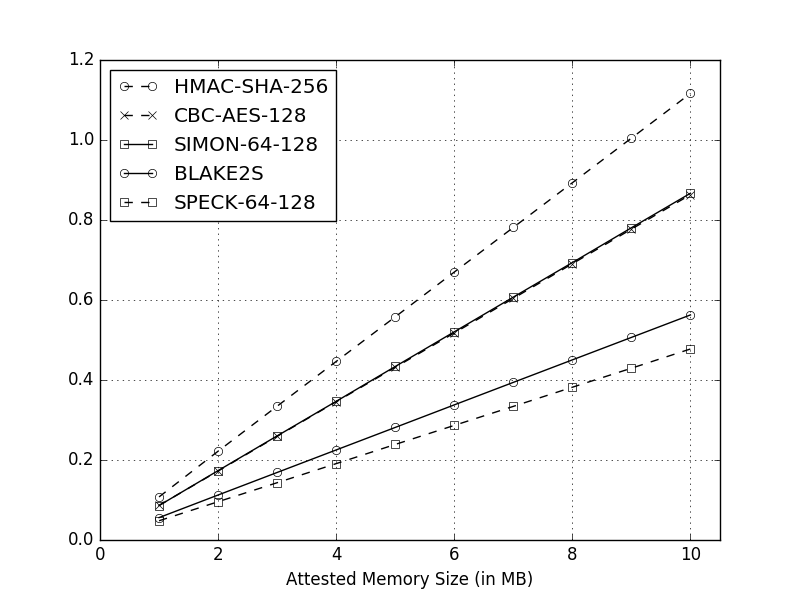}
		\caption{MacMem vs Mem Size}
		\label{partsize}
	\end{subfigure}
	~
	\begin{subfigure}[t]{0.3\textwidth}
		\centering
		\includegraphics[width=\linewidth]{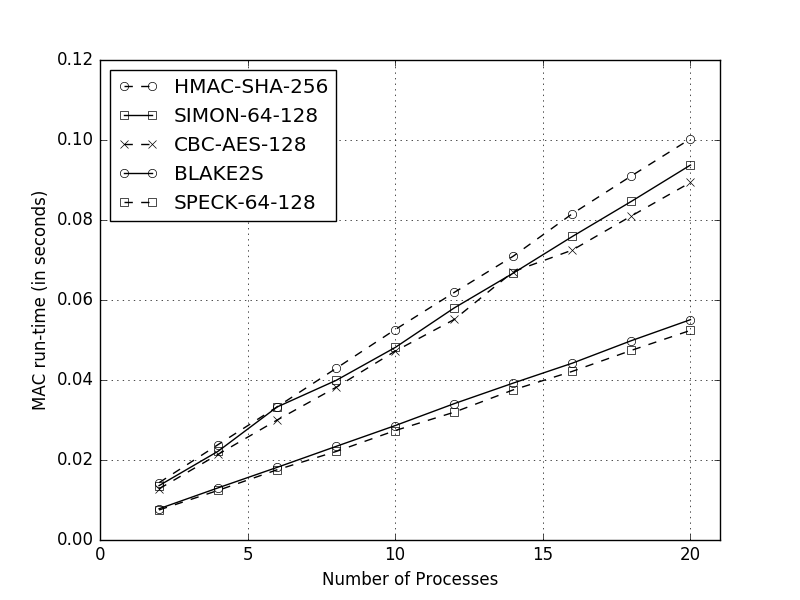}
		\caption{MacMem vs Num Processes}
		\label{numpart}
	\end{subfigure}

	\caption{Evaluation of HYDRA in SabreLite prototype}
\end{figure*}

In this section, we (informally) demonstrate how HYDRA satisfies the 
minimal set of requirements to realize secure RA 
(described in Section \ref{goals-assumptions}). The following are key features 
ensured in the design and implementation of \acro:

(1) the \sel\ microkernel is the first executable that is loaded in a \acro-based system 
upon boot/initialization. Correctness of this step is guaranteed by a 
ROM integrity check at boot time, e.g., HAB in the Sabre Lite case.

(2) The attestation process (\atp~)
\footnote{\atp\ is different from \att\ per Figure \ref{img_layout}. \atp\ is 
what is called ``initial process" in Figure \ref{img_layout} and it contains 
\att\  executable as a component.} 
is the first user-space process that 
is loaded into memory and is bootstrapped by
\sel . This is also guaranteed using a software integrity check 
step performed by \sel\ before spawning the initial process.

(3) \atp\ then starts with the highest scheduling priority and never
decreases its own priority value. This can be guaranteed by ensuring the code of 
\atp\ does not contain any system calls to decrease its priority.

(4) Any other subsequent process that is loaded on top of the \sel\ microkernel is spawned by 
$P_{Attest}$ and does not receive the highest scheduling priority. This can be 
ensured by inspecting the \atp\ code and ensuring that all invocations of 
other processes are with a lower priority value. Once a process is loaded with a certain priority, 
\sel~prevents it from increasing its priority value; this is formally verified and guaranteed 
in the \sel~implementation.

(5) The software executable and \key\ can only be mapped 
into the address space of \atp. This is guaranteed by ensuring that in the 
\atp~code no other process on initialization (performed in \atp~)
receives the capabilities to access said memory ranges.

(6) Virtual memory used by \atp\ cannot be used by
any other process; this includes any memory used for any computation touching 
the key, or related to other values computed using the key. This is 
formally verified and guaranteed in the \sel~implementation.

(7) Other processes cannot control or infer execution of \atp\ (protected by exclusive capability to TCB's \atp).

(8) Access control properties, i.e., authority confinement, integrity and
confidentiality, in \sel's binary are mathematically guaranteed 
by its formal verification.\\

Given the above features, the security properties in 
Section \ref{goals-assumptions} are satisfied because:

\textbf{Exclusive Access to \key:} 
(5), (6) and (8) guarantee that only \atp\ can have access to \key.

\textbf{No Leaks:} 
(6) and (8) ensures that intermediate values created by key-related computation inside
\atp\ cannot be leaked to or learned by other processes.

\textbf{Immutability:}
(1) and (2) implies that \acro~is initialized into the correct expected known initial states and that the correct binary
executable is securely loaded into RAM. (5) also prevents other 
processes from modifying that executable.

\textbf{Uninterruptability:}
(3) and (4) guarantees that other processes, always having a lower 
priority value compared to \atp, cannot interrupt the execution 
of \atp.

\textbf{Controlled Invocation:}
(7) ensures that the execution of \atp\ cannot be 
manipulated by other applications.

\section{Experimental Evaluation\label{evaluation}}
We present here performance evaluation of \acro\ using 
the Sabre Lite prototype. (Results of \acro\ 
on ODROID-XU4 are in Appendix \ref{appdx:odroid_perf}).
We conduct 
experiments to assess speed of, and overhead involved in, 
performing attestation using different types of keyed Message Authentication Code (MAC) functions,  
on various numbers of user-space processes and sizes of 
memory regions to be attested. 
We obtain the fastest performance using the Speck MAC; \acro\ can attest 
$10$MB in less than $500$msec in that case. 

\paragraph{Breakdown of Attestation Runtime}
Recall from Section \ref{designs}, that the attestation algorithm (Algorithm \ref{alg:mod_smart}) 
is composed of three operations.
$VerifyRequest$ (lines 3 to 9) is responsible for verifying an attestation request 
and whether it has been recently generated by an authorized verifier.
$RetrieveMem$ (line 11) maps memory regions from a target process to \atp's 
address space and returns a pointer to the mapped memory. $MacMem$ (lines 13 to 20) 
computes a cryptographic checksum (using \key) on the memory regions.

As shown in Table \ref{smart-time}, the runtime of $MacMem$ contributes the highest amount of 
the overall \att\ runtime: ~84\% of total time for attesting 1MB of memory and ~90\% 
for attesting 10 KB of memory on Sabre Lite;
whereas $RetrieveMem$ and $VerifyRequest$ together require less than 20\% of the overall time.

\paragraph{Performance of $RetrieveMem$ in \sel}
Another important factor affecting the performance of \acro\ is the runtime 
of $RetrieveMem$: the time \atp\ takes to map the 
attested memory regions to its own virtual address space. 
As expected, Figure \ref{retrieve_mem} illustrates the memory 
mapping runtime in \sel\ is linear in terms of mapped memory size.
In addition, we compare the runtime of $RetrieveMem$ and $MacMem$ on 
larger memory sizes. Figure \ref{retr_mac_ratio} illustrates that the 
runtime ratio of $RetrieveMem$ to various 
implementations of $MacMem$ is always less than 20\%.
This confirms that retrieving memory and mapping it to the address 
space account for only a small fraction of the total attestation time in \acro. 
This illustrates that whatever overhead \sel\ introduces when enforcing 
access control on memory is not significant and does not render \acro\ 
impractical.

\paragraph{Performance of $MacMem$ in \sel}
Since $MacMem$ is the biggest contributor to the runtime of our implementations, 
we explore various types of (keyed) cryptographic checksums and their 
performance on top of \sel. We compare the performance of five different MAC functions, 
namely, CBC-AES \cite{cbc-aes}, HMAC-SHA-256 \cite{sha2}, Simon and Speck \cite{simon-speck}, 
and BLAKE2S \cite{blake2}, on 1MB of data in the user-space of \sel.
The performance results in Figure \ref{macfn} illustrate that the runtime of 
MAC based on Speck-64-128
\footnote{Speck with 64-bit block size and 128-bit key size} 
and BLAKE2S in \sel\ are similar; and they are at least 33\% faster than other MAC 
functions when running on Sabre Lite.


\begin{savenotes}
\begin{table}[!h]
\centering
\caption{Performance Breakdown of Algorithm \ref{alg:mod_smart} on I.MX-SL @ 1GHz}
\label{smart-time}
\resizebox{\columnwidth}{!}{%
\begin{tabular}{ccccc}
\toprule 
    	\multirow{2}[3]{*}{Operations} & \multicolumn{2}{c}{1 MB of Memory} & \multicolumn{2}{c}{10 KB of Memory}\\

	\cmidrule(lr){2-3}
	\cmidrule(lr){4-5}

    	& Time in cycle & Proportion & Time in cycle & Proportion  \\
    	\midrule
    	$VerifyRequest$    & 3,992   & $<$0.01\%  &   3,961  & 0.75\% \\
    	$RetrieveMem$      & 8,731,858    & 15.11\%    &   47,136  & 8.93\% \\
    	$MacMem$
& 49,021,846   & 84.89\%    &   476,857  & 90.32\% \\
    	$Overall$     & 57,757,696    & 100\%    & 527,954  & 100\% \\
    	\bottomrule
\end{tabular}
}
\end{table}
\end{savenotes}

\paragraph{Performance of $MacMem$ vs Memory Sizes}
Another factor that affects $MacMem$'s performance 
is the size of memory regions to be attested.
We experiment by creating another process in the 
user-space and perform attestation on various sizes 
(ranging from 1MB to 10MB) of memory regions inside that 
process. As expected, the results of this experiment, illustrated 
in Figure \ref{partsize}, indicate that $MacMem$ performance 
is linear as a function of the attested memory sizes. This experiment 
also illustrates feasibility of performing attestation of 10MB of memory 
on top of \sel\ in \acro\ using a Speck-based MAC in less than half a second.


\paragraph{Performance on $MacMem$ vs Numbers of Processes}
This experiment answers the following question: How 
would an increase in number of processes affect the performance of \acro?
To answer it, we have the initial process spawn additional 
user-space processes (from 2 to 20 extra processes) and, then, perform
$MacMem$ on 100 KB memory data in one of those processes.
To ensure  fair scheduling of every process, we set priority 
of all processes (including the initial process) to the maximum priority. The result 
from Figure \ref{numpart}
indicates that the performance of $MacMem$ is reasonably 
linear as a function of the number of processes on Sabre Lite.

\section{Conclusions\label{conclusion}}

This paper presents the first hybrid Remote Attestation design, \acro, that 
leverages the formally verified \sel\ microkernel 
to instantiate memory and process isolation, and enforce access control 
to memory and other resources; such isolation and access control have 
been ensured through hardware in previous designs. 
We implement \acro\ on two commodity hardware platforms 
and demonstrate feasibility and practicality of 
hybrid RA schemes that significantly improve security in contemporary 
embedded and cyber-physical systems, and that can work on commodity 
hardware and require no modification to it while providing security guarantees 
that can be assured using automated formal methods. 

\bibliographystyle{abbrv}

\vskip 1 cm

This material is based on research sponsored by the Department of Homeland Security (DHS) Science and
Technology Directorate, Cyber Security Division (DHS S\&T/CSD) BAA HSHQDC-14-R-B00016, and the
Government of United Kingdom of Great Britain and the Government of Canada via contract number
D15PC00223. The views and conclusions contained herein are those of the authors and should not be interpreted as necessarily
representing the official policies or endorsements, either expressed or implied, of the Department of Homeland
Security, the U.S. Government, or the Government of United Kingdom of Great Britain and the Government of
Canada

\begin{appendices}

\begin{figure}[!h]
	
 \centering
    	\begin{subfigure}[t]{\linewidth}
		\centering
		\includegraphics[height=4cm, width=\linewidth]{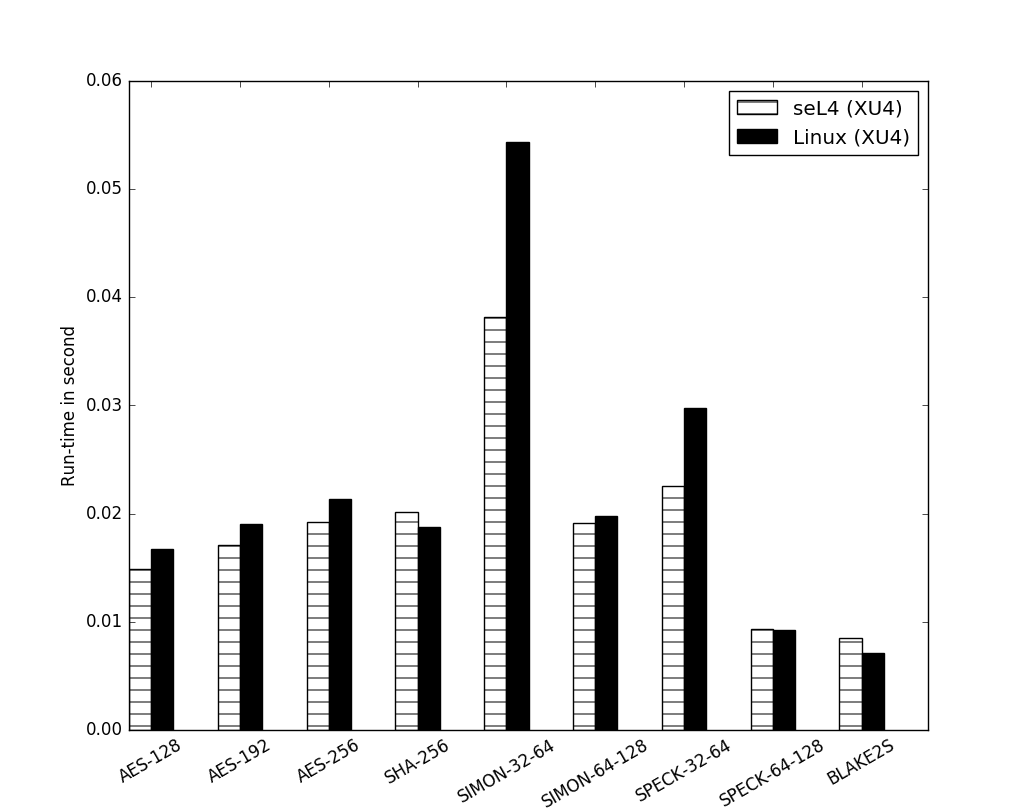}
		\caption{MAC Implementations}
		\label{macfn_odroid}
	\end{subfigure}
	~
    	\begin{subfigure}[t]{\linewidth}
		\centering
		\includegraphics[height=4cm, width=\linewidth]{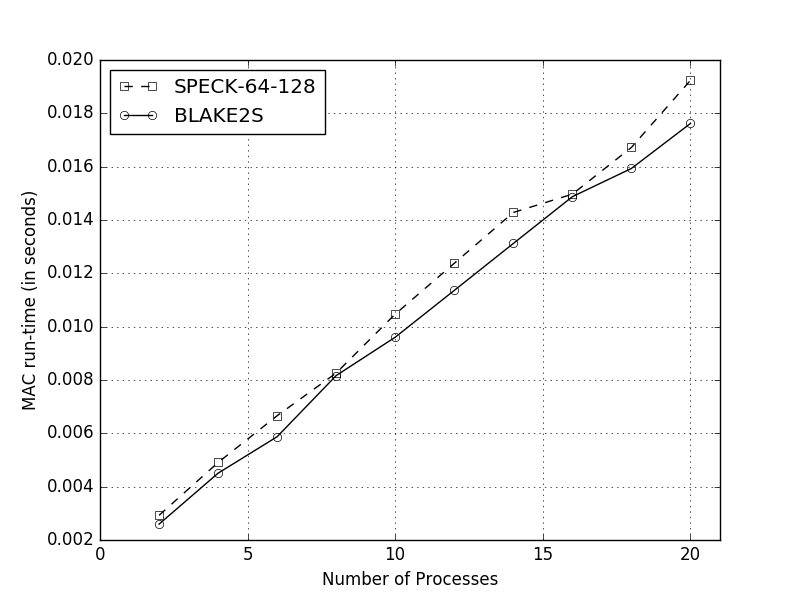}
		\caption{$MacMem$ vs Num Processes}
		\label{num_part_odroid}
	\end{subfigure}
	~
	\begin{subfigure}[t]{\linewidth}
		\centering
		\includegraphics[height=4cm, width=\linewidth]{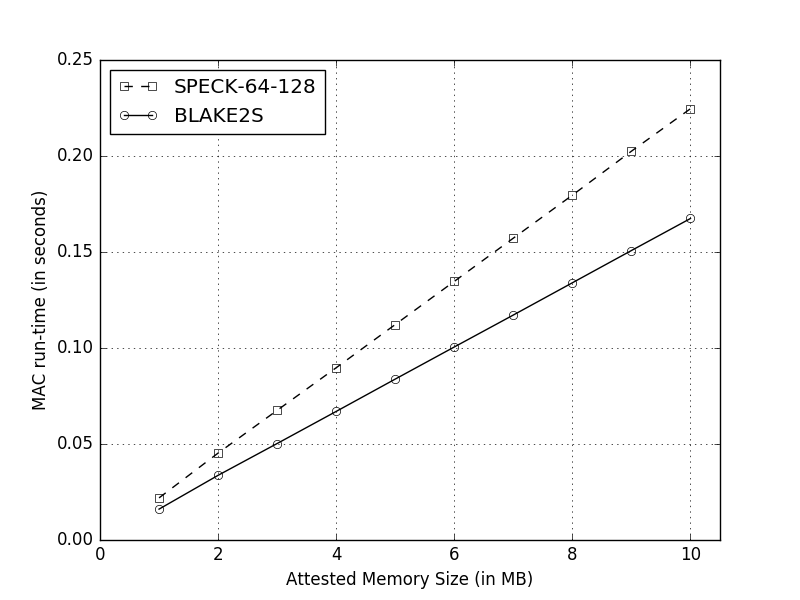}
		\caption{$MacMem$ vs Mem Size}
		\label{part_size_odroid}
	\end{subfigure}

	\caption{Evaluation of HYDRA in ODROID-XU4 prototype}
\end{figure}
\section{\acro's Performance in ODROID-XU4}\label{appdx:odroid_perf}

We also evaluate performance of \acro\ in ODROID-XU4 @ 2.1 GHz.
Despite not having an Ethernet driver, we evaluate the core component of \acro: $MacMem$. 
Unlike the results in Section \ref{evaluation}, BLAKE2S-based MAC
achieves the fastest performance for attesting 10MB on ODROID-XU4 platform.

\paragraph{MAC Performance in Linux vs in \sel}
Figure \ref{macfn_odroid} illustrates the performance comparison of various keyed MAC functions
on ODROID-XU4 hardware platform when running on Ubuntu 15.10
and \sel. The result emphasizes the feasibility of RA in \sel\ since the runtime of \sel-based RA
can be as fast as that of RA running on top of widely prominent Linux operating system.

\paragraph{MAC Performance on ODROIX-XU4}
The result in Section \ref{evaluation} and the one from above demonstrate that Speck- and
BLAKE2S-based MACs have the fastest attestation runtime in \sel. We conduct additional experiments on
those two MAC functions in ODROID-XU4 platform. Figure \ref{num_part_odroid} asserts the linear relationship
between number of processes and $MacMem$ runtime. In addition, the MAC runtime, shown in Figure \ref{part_size_odroid},
is also linear in memory sizes to be attested. Lastly, runtime of the BLAKE2S-based MAC function achieves less than 200 milliseconds
for attesting 10MB of memory regions in ODROID-XU4

\section{\sel's Proof Assumptions}\label{appdx:sel4_assump}
\sel's functional correctness proof is based on the following assumptions:
\begin{itemize}
\item \textbf{Assembly} - it assumes the correctness of ARM assembly code mainly for entry to and exit from the kernel 
and direct hardware accesses.
\item \textbf{Hardware} - it assumes hardware operates according to its specification and has not been tampered with.
\item \textbf{Hardware Management} - it assumes the correctness of the underlying hardware management,
including a translation look-aside buffer (TLB) and cache-flushing operations.
\item \textbf{Boot Code} - it assumes the correctness of code that boots the \sel\ microkernel into memory.
\item \textbf{Direct Memory Access (DMA)} - it assumes DMA is disabled or trusted.
\item \textbf{Side-channels} - it assumes there is no timing side-channels.
\end{itemize}

\section{Sample Code for Starting New Process}\label{appdx:new_process_code}

In our implementation, \atp\ creates a new empty process with the default configuration below:
\begin{lstlisting}%[caption=Config of New Process, label={child-process-config}]

int sel4utils_configure_process_custom(sel4utils_process_t *process, vka_t *vka, vspace_t *spawner_vspace, sel4utils_process_config_t config)
{
    int error;
    sel4utils_alloc_data_t * data = NULL;
    memset(process, 0, sizeof(sel4utils_process_t));
    seL4_CapData_t cspace_root_data = seL4_CapData_Guard_new(0, seL4_WordBits - config.one_level_cspace_size_bits);

    /* create a page directory */
    process->own_vspace = config.create_vspace;
    error = vka_alloc_vspace_root(vka, &process->pd);
    if (error) {
        goto error;
    }

    /* assign an asid pool */
    if (assign_asid_pool(config.asid_pool, process->pd.cptr) != seL4_NoError) {
        goto error;
    }

    /* create a cspace and copy its cap to the new process */
    process->own_cspace = config.create_cspace;
    if (create_cspace(vka, config.one_level_cspace_size_bits, process, cspace_root_data) != 0) {
        goto error;
    }

    /* create a fault endpoint and copy its cap to the new process */
    if (create_fault_endpoint(vka, process) != 0) {
        goto error;
    }

    /* create a vspace */
    sel4utils_get_vspace(spawner_vspace, &process->vspace, &process->data, vka, process->pd.cptr, sel4utils_allocated_object, (void *) process);

    /* finally elf load */
    process->entry_point = sel4utils_elf_load(&process->vspace, spawner_vspace, vka, vka, config.image_name);

    if (process->entry_point == NULL) {
        goto error;
    }

    /* create the thread */
    error = sel4utils_configure_thread(vka, spawner_vspace, &process->vspace, SEL4UTILS_ENDPOINT_SLOT, config.priority, process->cspace.cptr, cspace_root_data, &process->thread);

    if (error) {
        goto error;
    }

    return 0;

error:
    /*  clean up */
	...

    return -1;
}


int sel4utils_configure_thread_config(vka_t *vka, vspace_t *parent, vspace_t *alloc, sel4utils_thread_config_t config, sel4utils_thread_t *res)
{
    memset(res, 0, sizeof(sel4utils_thread_t));

    int error = vka_alloc_tcb(vka, &res->tcb);
    if (error == -1) {
        sel4utils_clean_up_thread(vka, alloc, res);
        return -1;
    }

    res->ipc_buffer_addr = (seL4_Word) vspace_new_ipc_buffer(alloc, &res->ipc_buffer);

    if (res->ipc_buffer_addr == 0) {
        return -1;
    }

    if (write_ipc_buffer_user_data(vka, parent, res->ipc_buffer, res->ipc_buffer_addr)) {
        return -1;
    }

    seL4_CapData_t null_cap_data = {{0}};
    error = seL4_TCB_Configure(res->tcb.cptr, config.fault_endpoint, config.priority, config.cspace, config.cspace_root_data, vspace_get_root(alloc), null_cap_data, res->ipc_buffer_addr, res->ipc_buffer);

    if (error != seL4_NoError) {
        sel4utils_clean_up_thread(vka, alloc, res);
        return -1;
    }

    res->stack_top = vspace_new_stack(alloc);

    if (res->stack_top == NULL) {
        sel4utils_clean_up_thread(vka, alloc, res);
        return -1;
    }

    return 0;
}

\end{lstlisting}

\end{appendices}

\end{document}